\documentstyle[aps,prl,amssymb,preprint]{revtex}  
\preprint{A. Einstein Institute preprint AEI-055 (1998), gr-qc/9801073}
\tighten

\draft

\font\SYMA=msam10 
\def\lrcorner{\hbox{\SYMA y}}

\def\lessthansquiggle{\raise.3ex\hbox{$<$\kern-.75em\lower1ex\hbox{$\sim$}}}

\def\greaterthansquiggle{\raise.3ex\hbox{$>$\kern-.75em\lower1ex\hbox{$\sim$}}}
\def\lessthansquiggle{\raise.3ex\hbox{$<$\kern-.75em\lower1ex\hbox{$\sim$}}}

\def\Reals{{\Bbb R}}

\def\Naturals{{\Bbb N}}

\newcommand{\dx}[1]{{\mbox{\rm d}#1}}

\newcommand{\di}{\displaystyle}
\newcommand{\rd}{{\rm d}}
\newcommand{\gr}[2]{\left|_{#1}^{#2} \right.}
\newcommand{\dr}[2]{{\partial{#1}\over \partial{#2}}}
\newcommand{\ddr}[3]{{\partial^2{#1}\over \partial{#2}\partial{#3}}}
\newcommand{\ddt}[1]{{{\rm d}{#1}\over {\rm d}t}}

\newcommand{\beq}{\begin{equation}}

\newcommand{\eeq}{\end{equation}}
\newcommand{\beqa}{\begin{eqnarray}}
\newcommand{\eeqa}{\end{eqnarray}}
\newcommand{\beqan}{\begin{eqnarray*}}
\newcommand{\eeqan}{\end{eqnarray*}}
\newcommand{\ba}{\begin{array}}
\newcommand{\ea}{\end{array}}

\newcommand{\vp}{\varphi}

\def\nz{\ifmmode {I\hskip -3pt N} \else {\hbox {$I\hskip -3pt N$}}\fi}
\def\zz{\ifmmode {Z\hskip -4.8pt Z} \else
       {\hbox {$Z\hskip -4.8pt Z$}}\fi}
\def\qz{\ifmmode {Q\hskip -5.0pt\vrule height6.0pt depth 0pt
       \hskip 6pt} \else {\hbox
       {$Q\hskip -5.0pt\vrule height6.0pt depth 0pt\hskip 6pt$}}\fi}
\def\rz{\ifmmode {I\hskip -3pt R} \else {\hbox {$I\hskip -3pt R$}}\fi}
\def\cz{\ifmmode {C\hskip -4.8pt\vrule height5.8pt\hskip 6.3pt} \else
       {\hbox {$C\hskip -4.8pt\vrule height5.8pt\hskip 6.3pt$}}\fi}
\def\au{{\setbox0=\hbox{\lower1.36775ex%
\hbox{''}\kern-.05em}\dp0=.36775ex\hskip0pt\box0}}
\def\ao{{}\kern-.10em\hbox{``}}

\newtheorem{Theorem}{Theorem}

\newcommand{\subjclass}[1]{}

\def\scri{\hbox{${\cal J}$\kern -.645em {\raise
      .57ex\hbox{$\scriptscriptstyle (\ $}}}}
\newcommand{\Scri}{\scri}

\newcommand{\eq}[1]{(\ref{#1})}

\newcommand{\be}{\begin{equation}}
\newcommand{\ee}{\end{equation}} \newcommand{\bea}{\begin{eqnarray}}
\newcommand{\eea}{\end{eqnarray}}
\newcommand{\beaa}{\begin{eqnarray*}}
\newcommand{\eeaa}{\end{eqnarray*}}

          \newcommand{\R}{{\Bbb R}}

\begin{document}                
\title{Uniqueness of the  mass in the radiating regime}

\author{Piotr T.\ Chru\'sciel\thanks{Supported in part by a 
Polish
    Research Council grant 
and by the Humboldt
    Foundation. The author acknowledges the hospitality of the A.Einstein
    Institute in Potsdam during part of work on this paper. E--mail:
Chrusciel@Univ-Tours.Fr}}
\address{D\'epartement de  Math\'ematiques, 
Facult\'e des Sciences, 
Universit\'e de Tours, 
Parc de Grandmont, F--37200 Tours, France}
\author{Jacek Jezierski\thanks{Partially supported by a grant from R\'egion
Centre. E--mail: jjacekj@fuw.edu.pl}}
\address{Department of Mathematical Methods in Physics, 
University of Warsaw, 
ul. Ho\.za 74, 00--682 Warszawa, Poland}
\author{Malcolm A.H.\ MacCallum\thanks{E--mail:
M.A.H.MacCallum@qmw.ac.uk}}
\address{
School of Mathematical Sciences, 
Queen Mary and Westfield College, 
University of London, 
Mile End Road, London E1 4NS, U.K.}
\maketitle
\begin{abstract}                
  The usual approaches to the definition of energy give an ambiguous
  result for the energy of fields in the radiating regime. We
  show that for a massless scalar field in Minkowski space--time the
  definition may be rendered unambiguous by adding the requirement
  that the energy cannot increase in retarded time. We present a
  similar theorem for the gravitational field, proved 
  elsewhere, which establishes that the Trautman--Bondi energy is the
  unique (up to a multiplicative factor) functional, within a natural
  class, which is monotonic in time for all solutions of the vacuum
  Einstein equations admitting a smooth ``piece'' of conformal null
  infinity \Scri.
\end{abstract}
\narrowtext

Consider a Lagrangian theory of fields $\phi^A$ defined on a
 manifold $M$ 
with a Lagrange function density
\be
\label{E.0}
{\cal L} = {\cal L}[\phi^A,\,\partial_\mu \phi^A,\,\ldots,\,
\partial_{\mu_1} \ldots \partial_{\mu_k} \phi^A]\ , \ee for some $k
\in \Naturals$, where $\partial_\mu$ denotes partial differentiation
with respect to $x^\mu$. Suppose further that there exists a function
$t$ on $M$ such that $M$ can be decomposed as $\R \times \Sigma$,
where $\Sigma\equiv\{t=0\}$ is a hypersurface in $M$ and the vector
$\partial/\partial t$ is tangent to the $\R$ factor. The proof of the
Noether theorem, as presented {\em e.g.\/} in [Section 10.1] of Ref.
\onlinecite{Sexl:Urbantke}, shows that the vector density
\begin{eqnarray}
&
E^\lambda =  - {\cal L} X^\lambda + X^\mu \sum_{\ell=0}^{k-1}
\phi^A{}_{,\alpha_1\ldots\alpha_\ell\mu}\qquad \qquad 
&\nonumber
\\
&
\qquad \times \sum_{j=0}^{k-\ell-1} (-1)^j
\partial_{\gamma_1}\ldots \partial_{\gamma_j}\left(
{{\partial {\cal L}} \over {\partial
 \phi^A{}_{,\lambda\alpha_1\ldots\alpha_{\ell}{\gamma_1}\ldots{\gamma_j}}}} 
\right)  
 & \label{E.1}
\end{eqnarray}
has vanishing divergence, ${E^\lambda }_{,\lambda}=0$, when the fields
$\phi_A$ are sufficiently smooth and satisfy the variational equations
associated with a sufficiently smooth ${\cal L}$ ({\em cf.\/} also
Ref.\ \onlinecite{WaldLee}). (This is in any case easily seen by
calculating the divergence of the right--hand--side of eq. \eq{E.1}.)
Here $ \phi^A{}_{,\alpha_1\ldots\alpha_\ell} =
\partial_{\alpha_1}\ldots\partial_{\alpha_\ell} \phi^A$, and
$X^\mu\partial_\mu=\partial_t$. In theories in which ${\cal L}$
depends only upon $\phi^A$ and its first derivatives, it is customary
to define the total energy associated with the hypersurface $\Sigma$
by the formula
\begin{equation}
\label{E.2}
E(\Sigma) = \int_\Sigma E^\lambda \dx S_\lambda,
\end{equation}
with $\dx S_\lambda = \partial_\lambda \,\lrcorner \,\dx x^0 \wedge
\ldots \dx x^3$, where $\lrcorner$ denotes contraction.
By extrapolation one can also use \eq{E.2} to define an ``energy'' for
higher order theories. 
Now it is well known that the addition to ${\cal L}$ of
a functional of the form
\be
\label{Yf}
\partial_\lambda(Y^\lambda[\phi^A,\,\partial_\alpha \phi^A,\,\ldots,\,
\partial_{\alpha_1} \ldots \partial_{\alpha_{k-1}} \phi^A])\ , \ee
where $k$ is as in \eq{E.0}, does not affect the field
equations\footnote{Here we adopt the standard point of view, that the
  field equations are obtained by requiring the action to be
  stationary with respect to all compactly supported variations ({\em
    cf. e.g.\/} Ref. \onlinecite{McCallumTaub} for a discussion of
  problems that might arise when this requirement is not enforced).}.
Such a change of the Lagrange function will change $E(\Sigma)$ by
a boundary integral (see, \emph{e.g.}, Ref.\ \onlinecite{CJM} for an
explicit formula for $\Delta E^{\mu\lambda}$):
\be
\label{E.4}
E(\Sigma) \ \longrightarrow \ \hat E(\Sigma) = E(\Sigma) +
\int_{\partial\Sigma}
\Delta
E^{\mu\lambda} dS_{\mu\lambda}\ , \ee where $ S_{\alpha\beta} =
\partial_\alpha\,\lrcorner\,\partial_\beta\,\lrcorner\, \dx x^0 \wedge
\ldots \wedge \dx x^3$.
If $\partial \Sigma$ is a ``sphere at infinity'' the integral over
$\partial \Sigma$ has of course to be understood by a limiting
process.  Unless the boundary conditions at $\partial \Sigma$ force
all such boundary integrals to give a zero contribution, if one wants
to define energy using this framework one has to have a criterion for
choosing a ``best'' functional, within the class of all functionals
obtainable in this way. The vanishing of such boundary integrals will
not occur in several cases of interest, including a massless scalar
field and general relativity in the radiation regime.

As an example, consider a scalar field $\phi$ in the Minkowski
space--time, with $\Sigma=\{t=0\}$.  Assume that $\phi$ satisfies the
rather strong fall--off conditions on $\Sigma$
\be
\label{E.5}
\partial_{\alpha_1}\ldots
\partial_{\alpha_j} \phi=o(r^{-2}),\quad 0\le j \le k-1\ , \ee where $k$
is the integer appearing in \eq{E.0}. In this case the boundary
integral in \eq{E.4} will vanish 
for all smooth
$Y^\mu$'s, as 
considered in eq.\ \eq{Yf}. This shows that  eq.\ \eq{E.2} leads to
a well--defined notion of energy on this space of fields (whatever the
Lagrange function ${\cal L}$), as long as the volume integral there
converges. (That will be the case if, {\em e.g.}, ${\cal L}$ has no
linear terms in $\phi$ and its derivatives.) 

Consider, next, a scalar field in Minkowski space--time, with
$\Sigma$ being a hyperboloid, $t=\sqrt{1+x^2+y^2+z^2}$. Suppose
further that in Minkowski coordinates 
${\cal L}={1\over 2}\eta^{\mu\nu}\partial_\mu\, \phi\partial_\nu\phi$,
so that the 
field equations read \be\label{E.6} \Box \phi = 0 \ .  \ee In that
case the boundary conditions \eq{E.5} would be incompatible with the
asymptotic behavior of those solutions of eq.\ \eq{E.6} which are
obtained by evolving compactly supported data on $\{t=0\}$ (see eq.\ 
\eq{F4new} below). Thus, even for scalar fields in Minkowski
space--time, a supplementary condition singling out a preferred
$E^\lambda$ is needed in the radiation regime.

Now for various field theories on the Minkowski background, including
the scalar field, one can impose some further conditions on
$E^\lambda$ which render it unique \cite{Fock,bicak:energy}, such as
Lorentz covariance, and dependence upon the first derivatives of the
field only. The extension of that analysis to the gravitational field
has been carried out in \cite{bicak:energy}, and it also leads to a
unique $E^\lambda$ (namely the one obtained from the so--called
``Einstein energy--momentum pseudo--tensor''), within the class of
objects considered. While this is certainly an interesting
observation, the restrictions imposed in that last paper on the
energy--momentum pseudo--tensor of the gravitational field are much more
restrictive than is desirable. Thus it seems of interest to find a
more natural criterion, which would encompass both general relativity
and field theories on a Minkowski background, and which would single out a
preferred expression for energy in the radiation regime. In this
letter we wish to point out that the requirement of \emph{monotonicity
  of energy in retarded time} allows one to single out an energy
expression in a unique way, within a natural class of ``energies''.
Let us start with the case of a massless scalar field in Minkowski
space--time. 
The variational formalism described above leads one to consider
functionals of the form 
\beqa \label{F1}
& H[\phi,t]=E(\Sigma_t)
+\int\limits_{\partial \Sigma_t}H^{\alpha\beta} \rd S_{\alpha\beta}\ ,
&
\\
&
E(\Sigma_t)= \int\limits_{\Sigma_t} {T^\mu}_\nu X^\nu dS_\mu,
\quad X^\nu\partial_\nu = \partial_t\ , \nonumber
&
\eeqa
where $H^{\alpha\beta}$ is a 
twice continuously differentiable 
function of $\phi(x)$, $\partial_{\alpha_1}\phi(x)$, $\ldots$,
$\partial_{\alpha_1} \ldots \partial_{\alpha_n} \phi(x)$, 
for some $n$. The indices $\alpha$ refer to Minkowski
coordinates, and in Minkowski coordinates the
$H^{\alpha\beta}$'s {\em depend upon the coordinates through the fields only}.
Here and throughout the $\Sigma_t$'s are unit hyperboloids in
Minkowski space--time:
$\Sigma_{t}=\{x^0=t+\sqrt{(x^1)^2+(x^2)^2+(x^3)^2}\}$, ${T^\mu}_\nu$
is the standard 
energy--momentum tensor for the scalar field
(with the normalization determined by eq.\ \eq{E.1}),  $\di
\int\limits_{\partial\Sigma_t} H^{\alpha\beta} \rd S_{\alpha\beta}$ is
understood as a limit as $R$ tends to infinity of integrals on
coordinate balls of radius $R$ included in $\Sigma_t$.

Before analyzing convergence of functionals \eq{F1} we need to specify
the class of fields $\phi$ of interest. Consider solutions of \eq{E.6}
which have smooth compactly supported initial data
on the hyperplane $\{x^0=0\}$, where $x^0$ is a standard Minkowski
coordinate. Using conformal covariance of eq.\ \eq{E.6} (\emph{cf.,
  e.g.,} arguments in Ref. \onlinecite{Waldbook}) it 
can be shown
that there will exist smooth 
functions $c (u,\theta,\phi)$ and $d(u,\theta,\phi)$ defined on
$(-\infty,\infty)\times S^2$ such that \be\label{F4new}
\phi(u,r,\theta,\phi)-{c(u,\theta,\phi)\over r}-
{d(u,\theta,\phi)\over r^2}=O( r^{-3}), \ee with $u=x^0-r$. Moreover
\eq{F4new} is preserved under differentiation in the obvious way. (The
hypothesis that the initial data are compactly supported is not
necessary, and is made only to avoid unnecessary technical
discussions.) In what follows we will only consider solutions of
\eq{E.6} satisfying \eq{F4new}. For our purposes it is important to
emphasize that \emph{given  arbitrary functions $c$ on, say $[u_0-1,
  u_0+1]\times S^2$ and $d_0$ on $S^2$ there exists a solution of
  the wave equation defined on Minkowski space--time such that
  \eq{F4new} holds, with}
\begin{equation}
  \label{eq:F4new.1}
  \frac{\partial d}{\partial u} = -\frac{1}{2} \Delta_2 c\ , \qquad
  d(u_0,\theta,\phi)=d_0(\theta,\phi)\ .
\end{equation}
Here $\Delta_2 $ denotes the Laplace operator on $S^2$ with the
standard round metric.  (Eq.\ \eq{eq:F4new.1} is obtained by inserting
the expansion \eq{F4new} in \eq{E.6}). This assertion is easily
proved using again the conformal covariance of eq.\ \eq{E.6}.  We
claim the following: 
\begin{Theorem} \label{T1}
Let $H$ be as described above and  suppose that \eq{F1} converges
for all  solutions  of the wave equation satisfying \eq{F4new}.
 If $\di\ddt{H} \leq 0$, then for all such
$\phi$'s  
\[ \int\limits_{\partial\Sigma_t} H^{\alpha\beta}\rd
S_{\alpha\beta} =0\ , \]
so that the numerical value of $H$ equals the standard canonical energy.
\end{Theorem}

{\sc Proof: 
  } We can Taylor expand $H^{\alpha\beta}$ at $\phi=0$ up to second
order to obtain \beqa\nonumber H^{\alpha\beta} & = &
H^{\alpha\beta}\Big|_{\phi=0}+
\sum_{0\leq |I|\leq k} \dr{H^{\alpha\beta}}{\phi_I} \Big| _{\phi=0}{\phi_I}\\
& & +\sum_{0\leq |I|,|J|\leq k} \ddr{H^{\alpha\beta}}{\phi_I}{\phi_J}
\Big| _{\phi=0}{} \phi_I\phi_J+r^{\alpha\beta}\ ,\label{F9} \eeqa
where we use the symbol $\phi_I$ to denote objects of the form
$\partial_{\alpha_1} \ldots \partial_{\alpha_\ell}\phi$, with
$|I|=|(\alpha_1,\ldots,\alpha_\ell)|=\ell$. By hypothesis
$H^{\alpha\beta}\gr{\phi=0}{}$ depends only upon the metric and its
derivatives, so in Minkowski coordinates the coefficients
$H^{\alpha\beta}\Big| _{\phi=0}$, $\dr{H^{\alpha\beta}}{\phi_I} \Big|
_{\phi=0}$ {\em etc.\/} in (\ref{F9}) are constants. By well known
properties of Taylor expansions and by (\ref{F4new}) we have
$r^{\alpha\beta}=o(r^{-2})$, so that $r^{\alpha\beta}$ will not
contribute to $H$ in the limit $r\to\infty$. 
In Minkowski coordinate systems (\ref{F4new}) can be
rewritten as 
\beqa\label{F10} &\dr{}{x^{\alpha_1}}
\ldots\dr{}{x^{\alpha_j}} \phi = {c^{(j)}\over r}n_{\alpha_1}\ldots
n_{\alpha_j}+{L_{\alpha_1\cdots\alpha_j}\over r^2} +O(r^{-3}),& \\
\label{F10.1} 
&\dr{}{x^{\alpha_1}} \ldots\dr{}{x^{\alpha_j}}{d
\over
  r^2}={d^{(j)}\over r^2}n_{\alpha_1}\ldots n_{\alpha_j}+O( r^{-3}) \ ,&
\eeqa where $c^{(m)}(t,r,\theta,\vp)=\dr{^m}{u^m}\left(
  c(u,\theta,\vp)\right) \gr{u=t-r}{}$, similarly for $d^{(m)}$, and
$n_\mu=(1, -{x^i\over r})$. Here $L_{\alpha_1\cdots\alpha_j}$ is a
linear functional of $c$ and its $u$ derivatives up to order
$j-1$.
Inserting \eq{F4new} in \eq{F9} and making use of \eq{F10}--\eq{F10.1}
might produce several terms which do not obviously converge, but those
have to cancel out or integrate out to zero by our hypothesis of
convergence of \eq{F1}. It then follows that \eq{F1} can be rewritten
as
\bea
\nonumber
H & = & E(\Sigma_t)
\\
& & +\int_{S^2}\hat
h[c,c^{(1)},\ldots,c^{(k)},d^{(1)},\ldots,d^{(k)},\theta,\phi]\, d^2\mu\ ,
\label{NE.1}
\eea
for some  functional 
$\hat h$, smooth in all its arguments, with $d^2\mu
=\sin\theta\rd\theta\rd\vp$. Moreover $\hat h$ is linear in 
$d$ and its $u$--derivatives.
Eq.\ \eq{eq:F4new.1} allows one to eliminate the $u$--derivatives of
$d$ in terms of derivatives of $c$, so that \eq{NE.1} can be rewritten as
\bea
\nonumber
H & = & E(\Sigma_t)
\\
\label{NE.2}
& &+\int_{S^2}\left(h[c,c^{(1)},\ldots,c^{(k)},\theta,\phi]
  +\alpha[d,\theta,\phi]\right)d^2\mu\ , \eea for some functionals
$h$ and $\alpha$, with $\alpha$ linear in $d$. The $u$--derivative of
\eq{NE.2} gives 
\begin{eqnarray*}
\frac{dH}{dt} & = & \int_{S^2}\left( -(c^{(1)})^2 + \frac{\delta h}{\delta
    c}c^{(1)} + \ldots \right.
\\
& & \qquad \qquad\left. +\frac{\delta h}{\delta c^{(k)}}c^{(k+1)} +
  \frac{\delta \alpha}{\delta d}d^{(1)} \right)d^2\mu\ , \end{eqnarray*}
Since $c^{(k+1)}$ is an arbitrary function on $S^2$ at fixed 
$c, \ldots, c^{(k)}$ and $d_0$, we can choose it so that
$\frac{dH}{dt}\le 0$ unless ${\delta h}/{\delta c^{(k)}}=0$, $k\ge 1$. A
suitable redefinition of $h$ leads to
\begin{eqnarray}
  & H(t) =
  E(\Sigma_t)+\int_{S^2}\left(h[c,\theta,\phi]+ \alpha[d,\theta,\phi]\right) 
d^2\mu\ , \nonumber \\
  \label{NE.3}
& \frac{dH}{dt}= \int_{S^2}\left( \left(-c^{(1)} + \frac{\delta h}{\delta
    c}\right)c^{(1)}  -  {1\over 2}\frac{\delta \alpha}{\delta
  d}\Delta_2c \right)d^2\mu\ ,& 
\end{eqnarray}
Consider now solutions of the wave equation with $c^{(1)}(u=u_0)=0$. In
this case \eq{NE.3} and arbitrariness of $c(u=u_0)$ imply that $dH/dt$
will be non--positive if and only if
$
\Delta_2({\delta \alpha}/{\delta
  d}) = 0 \ ,
$
which forces ${\delta \alpha}/{\delta
  d}$ to be a constant. We note that for any constant $a$ the integral
\be
\label{NE.5} \int_{S^2}a  d\,d^2\mu\ ,
\ee is a constant of motion (see also
\cite[Sect.\ 8.2]{JJAPP98}.
However, integrals of the form \eq{NE.5} cannot arise in the class of
functionals considered here. Indeed, the identity \eq{F10.1} shows
that all the terms which would give a non--vanishing contribution to
$H$ and which contain derivatives of $d$ contain at least one $u$
derivative of $d$. Then the only possible term which would contain $d$
would come from the term
\[
\int_{S^2}  \dr{H^{\alpha\beta}}{\phi} \Big|
_{\phi=0}{\phi}\,dS_{\alpha\beta} = 
\int_{S^2}  \dr{H^{rt}}{\phi} \Big| _{\phi=0}{(cr + d)}\,d^2\mu\ ,
\]
which for generic $c$ diverges when $r$ goes to infinity, unless
identically vanishing. We thus obtain 
${\delta \alpha}/{\delta d}= 0$. 
The right hand side
of eq.\ \eq{NE.3} can be made positive by choosing $c^{(1)}(u=u_0)=
{1\over 2} {\delta h}/{\delta c }$,
unless ${\delta h}/{\delta c}=0$, and our claim
follows. \hfill$\Box$

Let us now turn our attention to those gravitational fields which are
asymptotically Minkowskian in light--like directions.  An appropriate
mathematical framework here is that of space--like hypersurfaces which
intersect the future null infinity $\Scri^+$ in a compact
cross--section $K$. For such field configurations it  is widely
accepted that the ``correct'' definition of energy of a gravitating
system is that given by Freud \cite{F}, Trautman \cite{T,Tlectures2},
Bondi {\em et al.}\ \cite{BBM}, and Sachs \cite{Sachs}, which
henceforth will be called the Trautman--Bondi (TB) energy. 
There
have been various attempts to exhibit a privileged role of that
expression as compared with many alternative ones
(\cite{AshS,AshM,AshBR,Hecht:Nester,bicak:energy,%
Penrose:quasi-localmass,Penrose:Rindler:Ch9,%
Sachs:as,YorkBrown,TW,Geroch:WInicour}, to quote a few), but the
papers known to us have failed, for reasons sometimes closely related
to the ones described above, to give a completely unambiguous
prescription about how to define energy at \Scri. (We make some more
comments about that in \cite{CJM}, {\em cf.\/} also
\cite{Walker:Varenna}.) In a way rather similar to that for the case
of the scalar field described above, in  \cite{CJM} we show that 
{\em the TB
energy is, up to a multiplicative constant $\alpha \in \Reals$, the
{\em only functional of the gravitational field}, in a certain natural
class of functionals, which is {\em monotonic in time for all vacuum
  field configurations} which admit (a piece of) a smooth null
infinity $\Scri^+$}. More precisely, in \cite{CJM} we show the
following:

\begin{Theorem}
\label{T.2}
Let $H$ be a functional of the form
\begin{equation}
\label{E3.1}
H[g,\,u] = 
\int_{S^2(u)} H^{\alpha\beta}(g_{\mu\nu},\,
g_{\mu\nu,\sigma},\,\ldots,\,g_{\mu\nu,\sigma_1\ldots\sigma_k}) \,\dx
S_{\alpha\beta},
\end{equation}
where the $H^{\alpha\beta}$ are twice differentiable functions of
their arguments, and the integral over ${S^2(u)}$ is understood as a
limit as $\rho$ goes to infinity of integrals over the spheres
$t=u+\rho,r=\rho$. Suppose that $H$ is monotonic in $u$ for all vacuum
metrics $g_{\mu\nu}$ for which $H$ is finite, provided that
$g_{\mu\nu}$ satisfies
\begin{eqnarray}&
{g_{\mu\nu} = \eta_{\mu\nu} +
\frac{h^1_{\mu\nu}(u,\,\theta,\,\phi)}{r} +
\frac{h^2_{\mu\nu}(u,\,\theta,\,\phi)}{r^2} + o(r^{-2})}\ ,
& \nonumber \\
&{\partial_{\sigma_1} \ldots \partial_{\sigma_i}( g_{\mu\nu} -
\frac{h^1_{\mu\nu}(u,\,\theta,\,\phi)}{r} -
\frac{h^2_{\mu\nu}(u,\,\theta,\,\phi)}{r^2}) = o(r^{-2}),\quad
}
\label{E.3.1}
\end{eqnarray}
with $ 1\leq i \leq k$, for some $C^k$ functions
$h^a_{\mu\nu}(u,\,\theta,\,\phi)$, $a=1,2$. 
If $H$ is invariant under passive BMS 
super-translations, then the numerical value of $H$ equals (up to a
proportionality constant) the Trautman--Bondi mass.
\end{Theorem}

Some comments are in order. First, the volume integral $E(\Sigma_t)$ which
was present in \eq{F1} does not occur in \eq{E3.1}, because the
Trautman--Bondi mass is itself a boundary integral. Next, Theorem
\ref{T.2} imposes the further requirement of {\em passive BMS
  invariance}, which did not occur in the scalar field case. This
requirement arises as follows: recall that the coordinate systems in
which the metric satisfies \eq{E.3.1} are, roughly speaking, defined
only modulo BMS transformations. Then the requirement of {\em passive
  BMS invariance} is the rather reasonable requirement that the
concept of energy be independent of the coordinate system chosen to
measure this energy. We note that we believe that the requirement of
monotonicity  forces the energy to be invariant under (passive)
super--translations, but we have not succeeded in proving this so far.

The proof of Theorem \ref{T.2} is similar to the proof for the scalar
field presented here, but technically rather more involved. A key
ingredient of the proof is the Friedrich--Kannar \cite{F1,F2,K}
construction of space--times ``having a piece of $\Scri$''.

It is natural to ask why the Newman--Penrose constants of motion
\cite{NP2}, or the logarithmic constants of motion of \cite{ChMS}, do
not occur in our results of \cite{CJM}. These quantities are excluded
by the hypothesis that the boundary integrand $H^{\alpha\beta}$ which
appears in the integrals we consider depends on the coordinates only
through the fields. The Newman--Penrose constants could be obtained as
integrals of the form (\ref{E.1}) ({\em cf.\ e.g.\/} \cite{vdB}) if
explicit $r^{2}$ factors were allowed in $H^{\alpha\beta}$. Similarly
logarithmic constants could occur as integrals of the form (\ref{E.1})
if explicit $1/\ln r$ or $r^{+i}\ln^{-j} r$ factors were allowed
there.

Let us finally mention that one can set up a Hamiltonian framework in
a phase space which consists of Cauchy data on hyperboloids together
with values of the fields on appropriate parts of Scri to describe the
dynamics in the radiation regime \cite{CJK}. Unsurprisingly, the
Hamiltonians one obtains in such a formalism are again not unique, but
the non--uniqueness can be controlled in a very precise way. The
Trautman--Bondi mass turns out to be a Hamiltonian, and an appropriate
version of the uniqueness Theorem \ref{T.2} can be used to single out
the TB mass amongst the family of all possible Hamiltonians. In the
Hamiltonian framework the freedom of multiplying the functional by a
constant disappears.


\end{document}